\documentclass[showpacs,aps,prb,letterpaper,twocolumn,nofootinbib,raggedbottom,nobalancelastpage]{revtex4}
\usepackage{graphicx,bm}
\usepackage{amsmath}
\usepackage{color}

\begin{document}
\title{Generation of entangled photon-pairs from
a single quantum dot embedded in  a planar photonic-crystal cavity}
\date{\today}
\author{P. K. Pathak and S. Hughes}
\address{Department of Physics,
Queen's University, Kingston, ON K7L 3N6, Canada}
\begin{abstract}
We present a formal theory of single quantum-dot coupling
to a planar photonic crystal that supports
quasi-degenerate cavity modes, and use this theory to
describe, and optimize, entangled-photon-pair generation via the biexciton-exciton
cascade.
In the generated photon pairs, either both photons are spontaneously emitted from the dot, or one photon is emitted from
the biexciton spontaneously and the other is emitted via the leaky-cavity mode.
In the strong-coupling regime,
the generated photon pairs can be maximally entangled, in qualitative agreement with
the  simple dressed-state predictions of Johne {\em et al.} [Phys. Rev. Lett. vol.\,100, 240404 (2008)].
We derive useful and physically-intuitive analytical formulas for the spectrum of the emitted photon pairs in the presence of exciton and biexciton broadening, which is necessary to connect to experiments, and  demonstrate the
clear failure of using a dressed-state approach. We also present a method for
calculating and optimizing the entanglement between the emitted photons, which
can account for post-sample spectral filtering. Pronounced entanglement values
of greater than 80\% are demonstrated using experimentally achievable
parameters, even without spectral filtering.

\end{abstract}
\pacs{03.65.Ud, 03.67.Mn, 42.50.Dv}
\maketitle

\section{Introduction}
A source of polarization-entangled photon pairs has wide uses in quantum optics \cite{optics,optics2,pathak},
leading to applications such as quantum computation \cite{qcomp,qcomp2}, quantum information processing \cite{qinfo,qinfo2}, quantum cryptography \cite{crypto}, and quantum metrology \cite{metrology}. Most of the  experiments
demonstrated to date employ entangled photons generated by parametric down convertersion (PDC) \cite{mandelwolf,kwait}. A PDC is a ``heralded'' source of entangled photons in which the number of generated photon pairs is probabilistic. However, in many experiments, particularly in  quantum information processing \cite{scalable}, a deterministic scalable source of entangled photons is essential. Recently, there
has been considerable interest in developing  an all-solid-state ``on demand'' source of entangled photon pairs using single quantum dots  (QDs) \cite{akopian,stevenson,johne}. In  QDs, entangled photon pairs can be generated in a biexciton cascade decay via exciton states of angular momenta $+1$ and $-1$; single QDs are particularly appealing
since they are fixed in place, scalable, and have long coherence times.
 However, a major difficulty for implementing these schemes is the naturally occurring anisotropic energy difference
 between the exciton states of different angular momentum \cite{anisotropy}. Specifically, a small anisotropic energy difference, can make the emitted $x-$polarized and $y-$polarized photon pairs distinguishable, and thus the entanglement between the photons is largely wiped out. There have been a few proposals to overcome this problem, e.g., by spectrally filtering  indistinguishable photon pairs \cite{akopian}, by applying external fields to make the exciton states degenerate \cite{stevenson},
and suppressing the biexciton
binding energy in combination with time
reordering \cite{avronPRL08,ReimerARCH07},
but  these techniques have their own set of problems and are far from
optimal.

Recently, Johne {\em et al.}\,\cite{johne} proposed an interesting cavity-QED scheme in the strong coupling regime,
where the exciton states become dressed with the cavity field and form polariton states \cite{polariton}.
At the same time, a number of experimental groups
have now demonstrated the strong-coupling regime
using single QDs integrated with
planar photonic-crystal cavities  \cite{photocavity1,photocavity2,Hennessy}.
These emerging ``on-chip'' cavity structures form an important breakthrough in the search
for  creating scalable sources of photons using single QDs, and much excitement is envisioned. However,
the lack of appropriate theoretical descriptions becomes very challenging and the development of new medium-dependent
models are required to properly describe the  light-matter interactions and photon
wave functions.

In quantum material systems such as solids, the interaction with the environment is inevitable. The biexcitons, excitons and cavity modes interact with their phonon and thermal reservoirs \cite{dephasing2,dephasing,frank,new_tejedor}, which can have a substantial influence on the wavefunction of the emitted photon pairs. In the biexciton decay, the entanglement depends on the ``indistinguishability'' between $x-$polarized and $y-$polarized photon pairs, i.e. the overlap of their wavefunctions. Therefore, the precise form of the wave functions of the emitted  photon pairs is ultimately required. Here we present rigorous, and
physically-intuitive, analytical expressions for the wavefunction of the emitted photon pairs in the biexciton-exciton cascade decay using the Weisskopf-Wigner approximation for coupling to the environment. Extending previous
approaches \cite{johne}, we necessarily consider finite exciton and biexciton level broadenings and the damping of the leaky cavity mode. We further apply a method for optimizing  the entanglement using a simple spectral filter \cite{akopian}, and find impressive entanglement values even with realistic parameters and a sizable
anisotropic energy exchange.

\section{Theory}
We consider a QD embedded in a photonic crystal cavity having two orthogonal polarization modes of frequency $\omega_c^x$ and $\omega_c^y$, which can be realized and tuned experimentally using
e-beam lithography and, for example, AFM oxidization techniques \cite{HennessyAPL06}. The exciton states, $|x\rangle$ and $|y\rangle$, have an anisotropic-exchange energy difference $\delta_{\rm x}$. The cavity modes are coupled with the exciton to ground-state transition,
but  spectrally decoupled
from the biexciton state because of the relatively large biexciton binding energy, $\Delta_{xx}\gg$ cavity coupling.
The schematic arrangement of the system is shown in Fig.\,1.
For simplicity, we consider the emission of $x-$polarized photon pair, but the formalism and results
can easily be applied to $y-$polarized photons as well.
The Hamiltonian for the emission of $x-$polarized photon pair, in the interaction picture, can be  written as
\begin{eqnarray}
&&\!\!\!H_{I}(t) =\hbar\left[g|x\rangle\langle g|\hat a_c e^{i\Delta_c^x t}+\sum_{k\neq c}\Omega_{uk}|u\rangle\langle x|\hat a_k e^{i(\omega_{ux}-\omega_k)t}\right.\nonumber\\
&&\!\!\!\!\!\!\left.+\sum_{l\neq c}\Omega_{gl}|x\rangle\langle g|\hat a_l e^{i(\omega_{x}-\omega_l)t}
\!+\!\sum_{m\neq c}\Omega_{cm} \hat a_c^{\dagger} \hat a_m e^{i(\omega^x_c-\omega_m)t}\right]\!\!, \  \ \  \ \
\end{eqnarray}
plus a Hermitian conjugate term,
where $\omega_{ux}=\omega_u-\omega_x$, $\Delta_c^x=\omega_x-\omega^x_c,$
and $\hat a_i$ is the field operators with $\hat a_c$ the cavity mode operator.
Here, $\Omega_{uk}$, $\Omega_{gl}$, $\Omega_{cm}$ represent the couplings to the environment from
the biexciton, exciton and cavity mode; $g$ is the coupling between the
exciton and cavity mode; and $\omega_k$, $\omega_l$, $\omega_m$, $\omega_u$, and $\omega_x$ are the frequency of the photon emitted from the biexciton and exciton, the frequency of photon leaked from cavity, and the frequency of the biexciton and exciton, respectively.
We consider a system that is pumped in such a way as to have
an initially-excited biexciton, with no photons inside the cavity, thus
the state of the system at any time $t$ can be written as
\begin{eqnarray}
&&\!\!\!\!\!\!\!\!\!|\psi(t)\rangle\!\!=\!\!c_1(t)|u,0\rangle\!+\!\!\sum_kc_{2k}(t)|x,0\rangle|1_k\rangle\!\!+\!\!\sum_kc_{3k}(t)|g,1\rangle|1_k\rangle\nonumber\\
&&\!\!\!+\sum_{k,l}c_{4kl}(t)|g,0\rangle|1_k,1_l\rangle+\sum_{k,m}c_{5km}(t)|g,0\rangle|1_k\rangle|1_m\rangle. \ \
\end{eqnarray}
The different terms in the state vector $|\psi\rangle$ represent, respectively, the dot is in the biexciton state with zero photons in the cavity, the dot is in the exciton state after radiating one photon, the dot is in ground state with one photon in cavity mode, the dot is in the ground state after radiating two photons, and the dot is in ground state after one photon is radiated from the biexciton and the other is leaked from the cavity mode.

\begin{figure}[t]
\centering
\includegraphics[width=3.4in]{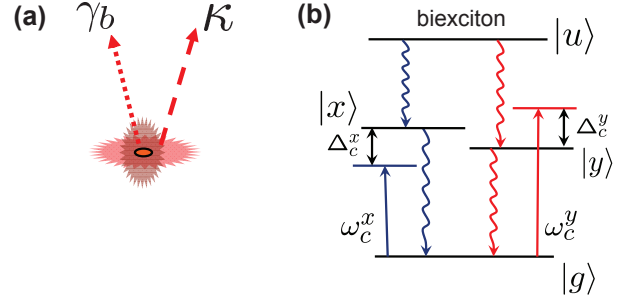}
\vspace{-0.3cm}
\caption{Schematic of the planar photonic crystal containing a single QD (a), and
the resulting energy level diagram for cavity-QED assisted generation of entangled photons in the biexciton-exciton
cascade decay (b). The biexciton state $|u\rangle$ decays to the ground state $|g\rangle$ via intermediate exciton state $|x\rangle$ or $|y\rangle$, emitting $x-$polarized or $y-$polarized photon pair. The $x-$polarized and $y-$polarized cavity modes are coupled with $|x\rangle\rightarrow|g\rangle$ and $|y\rangle\rightarrow|g\rangle$ transition, respectively. The
vertical decays are caused by leaky cavity-mode decay ($\kappa$) and by background
radiation modes ($\gamma_b$) above the photonic-crystal slab light line.} \label{fig1}
\vspace{-0.1cm}
\end{figure}

By using the Schr\"odinger equation, the equation of motion for the probability amplitudes are
\begin{eqnarray}
\label{c1}
\dot{c}_1(t)&=&-i\sum_k\Omega_{uk}c_{2k}(t)e^{i(\omega_{ux}-\omega_k)t} ,\\
\label{c2}
\dot{c}_{2k}(t)&=&-i\Omega^*_{uk}c_1e^{-i(\omega_{ux}-\omega_k)t}-igc_{3k}(t)e^{i\Delta_c^x t}\nonumber\\&&-i\sum_l\Omega_{gl}c_{4kl}(t)e^{i(\omega_x-\omega_l)t},\\
\label{c3}
\dot{c}_{3k}(t)&=&-igc_{2k}(t)e^{-i\Delta t}-i\sum_{m}\Omega_{cm}c_{5km}(t)e^{i(\omega-\omega_m)t}, \ \ \ \ \\
\label{c4}
\dot{c}_{4kl}(t)&=&-i\Omega^*_{gl}c_{2k}(t)e^{-i(\omega_x-\omega_l)t},\\
\dot{c}_{5km}(t)&=&-i\Omega^*_{cm}c_{3k}(t)e^{-i(\omega-\omega_m)t}.
\end{eqnarray}
Applying the Weisskopf-Wigner approximation \cite{wwa},
then  Eqs. (\ref{c1})- (\ref{c3}) simplify to
\begin{eqnarray}
\dot{c}_1(t)&\!=\!\!&-\gamma_1c_1(t) ,\\
\dot{c}_{2k}(t)&\!\!=\!&-i\Omega^*_{uk}c_1(t)e^{-i(\omega_{ux}-\omega_k)t}-igc_{3k}(t)e^{i\Delta_c^x t}
-\gamma_2c_{2k}(t), \ \ \ \ \\
\dot{c}_{3k}(t)&\!=\!\!&-igc_{2k}(t)e^{-i\Delta_c^x t}-\kappa c_{3k}(t),
\label{nc3}
\end{eqnarray}
where $\kappa=\pi|\Omega_{cm}|^2$ is the half width of the cavity mode and $\gamma_1$, $\gamma_2$ are
the half widths of the  biexciton and exciton levels, respectively.
We note that  $\gamma_1$ and $\gamma_2$ can include
both  radiative and nonradiative broadening, and for QDs, $\gamma_1\approx2\gamma_2$. Moreover, the radiative half width of biexciton will be sum of its spontaneous decay rates in the exciton states $|x\rangle$ and $|y\rangle$; if the decay rate of the biexciton in $|x\rangle$ and $|y\rangle$ are equal, the radiative half width of biexciton will be $2\pi|\Omega_{uk}|^2$. The radiative half width of the exciton $|x\rangle$ is given by $\gamma_b=\pi|\Omega_{gl}|^2$.
We next solve Eqs.(\ref{c4})-(\ref{nc3}) for $c_{4kl}$ and $c_{5km}$ using the Laplace transform method.
The probability amplitudes for two-photon  emission in long time limit are given by
\begin{eqnarray}
\label{c4kl}
c_{4kl}(\infty)&=&\frac{\Omega^*_{uk}}
{(\omega_k+\omega_l-\omega_u+i\gamma_1)} \times \nonumber\\
&& \frac{\Omega^*_{gl}(\omega_l-\omega_c^x+i\kappa)}{(\omega_l-\omega_x+ig_+)(\omega_l-\omega_x+ig_-)},\\
\label{c5km}
c_{5km}(\infty)&=&\frac{\Omega^*_{uk}}
{(\omega_k+\omega_m-\omega_u+i\gamma_1)} \times \nonumber\\
&& \frac{g\Omega^*_{cm}}{(\omega_m-\omega_x+ig_+)(\omega_m-\omega_x+ig_-)},
\label{gpm}
\end{eqnarray}
where $g_{\pm}=0.5(\kappa+\gamma_2-i\Delta_c^x\pm i\sqrt{4g^2-(\kappa-\gamma_2-i\Delta_c^x)^2})$.
In the case if there is no cavity coupling, namely $g=0$, the photons will be emitted spontaneously from the dot,
and we obtain a limiting form
\begin{equation}
c_{4kl}(\infty)=\frac{\Omega^*_{uk}\Omega^*_{gl}}{(\omega_k+\omega_l-\omega_u+i\gamma_1)(\omega_l-\omega_x+i\gamma_2)},
\end{equation}
which is the two-photon emission probability amplitude from a cascade
in free space, in agreement with
results of Akopian {\em et al.} \cite{akopian}. Thus, the influence of the cavity is determined by $g_{\pm}$, as one might expect. Next,
the optical spectrum of the {\em spontaneously} emitted photon-pair spontaneously, via {\em radiation} modes
above the photon crystal light line, is given by $S_r(\omega_k,\omega_l)=|c_{4kl}(\infty)|^2$, where
\begin{eqnarray}
 S_{r}(\omega_k,\omega_l)&=&\frac{|\Omega_{uk}|^2}{[(\omega_k+\omega_l-\omega_u)^2+\gamma_1^2]}\times\nonumber\\
 &&\frac{|\Omega_{gl}|^2|(\omega_l-\omega_c^x+i\kappa)|^2}
 {|(\omega_l-\omega_x+ig_+)(\omega_l-\omega_x+ig_-)|^2}. \
\end{eqnarray}
Similarly, the spectrum of the photon-pair with one photon emitted spontaneously from the biexciton and the other
photon emitted via the leaky-cavity mode
(cf.\,Fig\,(1)) is $S_{c}(\omega_k,\omega_m)=|c_{5km}(\infty)|^2$, where
\begin{eqnarray}
 S_c(\omega_k,\omega_m)&=&\frac{|\Omega_{uk}|^2}{[(\omega_k+\omega_m-\omega_u)^2+\gamma_1^2]}\times\nonumber\\
 &&\frac{g^2|\Omega_{cm}|^2}{|(\omega_m-\omega_x+ig_+)(\omega_m-\omega_x+ig_-)|^2}. \ \ \ \ \ \
\end{eqnarray}
The spectral functions $S_{r}(\omega_k,\omega_l)$ and $S_{c}(\omega_k,\omega_m)$
represent the joint probability distribution, and thus the integration over the one frequency
variable gives the spectrum at the other frequency. For example, the spectrum of the photon coming from the spontaneous decay of the exciton decay will be $S_r(\omega_l)=\int_{-\infty}^{\infty} S_r(\omega_k,\omega_l)\,d\omega_k$, and the spectrum of photon emitted via cavity mode is  $S_c(\omega_m)=\int_{-\infty}^{\infty} S_c(\omega_k,\omega_m)\,d\omega_k$, yielding
\begin{eqnarray}
\label{rad}
S_r(\omega_l)&=&\frac{|\Omega_{gl}|^2|(\omega_l-\omega_c^x+i\kappa)|^2}{|(\omega_l-\omega_x+ig_+)(\omega_l-\omega_x+ig_-)|^2} ,\\
\label{cav}
S_c(\omega_m)&=&\frac{ g^2|\Omega_{cm}|^2}{|(\omega_m-\omega_x+ig_+)(\omega_m-\omega_x+ig_-)|^2}, \
\end{eqnarray}
which is similar to the radiation-mode and cavity-mode emitted spectra
reported by Cui and Raymer \cite{Raymer:PRA06}, and by Yao and Hughes \cite{usPRLSubmit}.
From Eqs.(\ref{rad}) and (\ref{cav}), the photon emitted from the exciton decay (second emitted photon) has a
two-peak spectrum; these spectral peaks appear at the frequencies, $\frac{1}{2}\left(\omega_x+\omega_c^x\pm\delta\omega\right)$, where $\delta\omega\approx\sqrt{4g^2+\Delta_c^{x2}-(\kappa-\gamma_2)^2}$ is the splitting between the peaks.
In a dressed-state picture, these spectral peaks correspond to the two polariton states in the strong cavity regime,
 $g>>(\kappa,\gamma_2)$ \cite{johne}.

From the above discussion, the state of the ``photon pair'' emitted from both $|x\rangle$-exciton and $|y\rangle$-exciton branches is given by
\begin{eqnarray}
 &&\!\!\!\!\!\!\!\!|\psi(\infty)\rangle\!=\!\!\sum_{k,l}c_{4kl}(\infty)|1_k,1_l\rangle_x|{0}\rangle_x+\sum_{k,m}c_{5km}(\infty)|1_k\rangle_x|1_m\rangle_x\nonumber\\
 &&\!\!+\sum_{k,l}d_{4kl}(\infty)|1_k,1_l\rangle_y|{0}\rangle_y+\sum_{k,m}d_{5km}(\infty)|1_k\rangle_y|1_m\rangle_y, \ \ \
\label{state}
\end{eqnarray}
where in each term the first ket represents the combined state of the biexciton and the exciton reservoirs, the second ket represents the state of the cavity reservoir,
and the ket suffix  labels the polarization. The coefficients $c_{ijk}(\infty)$ are given by Eqs.(\ref{c4kl})-(\ref{gpm}).
For the same cavity coupling $g$, the coefficients, $d_{ijk}$, are given by the Eqs.(\ref{c4kl})-(\ref{gpm}) after replacing $\omega_x$, $\omega_c^x$, $\Delta_c^{x}$ with $\omega_y$, $\omega_c^y$, and $\Delta_c^y=\omega_y-\omega_c^y$, respectively.

\begin{figure}[b!]
\centerline{
\includegraphics[width=3.5in, height=3in]{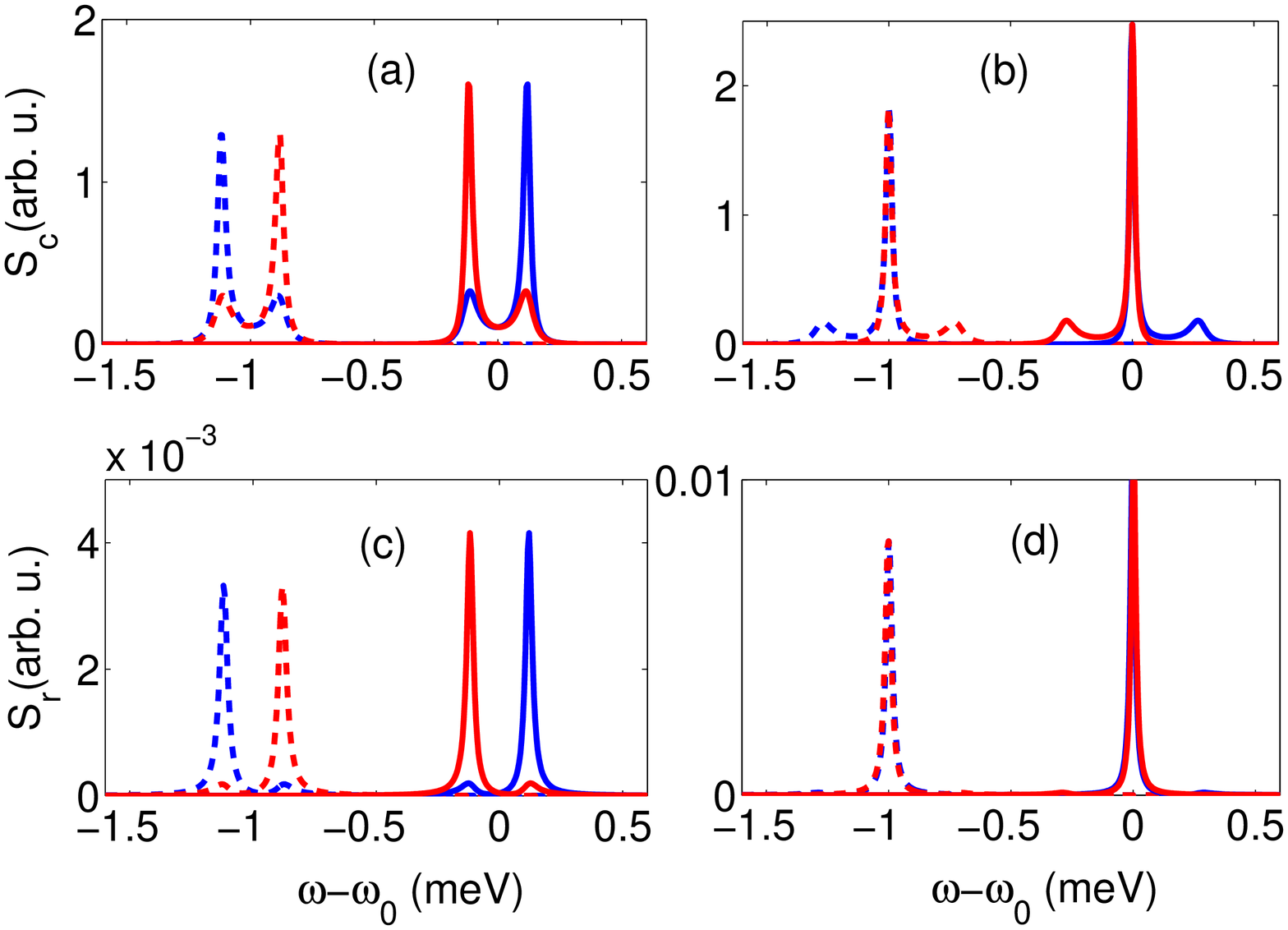}}
\vspace{-0.3cm}
\caption{The spectrum of the generated photon pair in the biexciton-exciton cascade decay, for $\delta_x=0.1$\,meV, $\Delta_{xx}=1.0$\,meV, $\gamma_1=2\gamma_2=0.004\,$meV, $\gamma_b=\pi|\Omega_{uk}|^2=\pi|\Omega_{gl}|^2=0.05\,\mu$eV \cite{Hennessy}, $\kappa=0.05$\,meV, $g=0.11\,$meV, and $\omega_0=0.5(\omega_x+\omega_y)$. In (a) and (b), one photon is emitted from biexciton decay and the other is leaked via cavity mode, and in (c) and (d), both photons are radiated from the biexciton and exciton states via spontaneous decay. The other parameters are as follows: for (a) and (c), $\Delta_c^x=-\Delta_c^y=\delta_x$, and for (b) and (d), $\Delta_c^x=-\Delta_c^y=-0.175\,$meV. The $x-$polarized photons are shown in blue and $y-$polarized
are shown in red; the solid curves are for photons generated in the exciton decay and dotted curves are for photons generated in the biexciton decay.}
\vspace{-0.1cm}
\label{fig2}
\end{figure}

\section{Results and Optimizing the Entanglement}
There are two possible decay channels for generating a photon pair.
In Figs.\,2(a) and (b), we show the spectrum of the photon pair, when one photon is emitted from biexciton decay and
the second is emitted (leaked) via the cavity mode. The spectra of photon pairs emitted in the biexciton and exciton radiative decay areshown in Figs.\,2(c) and 2(d). Depending on the  detunings between the frequency of the cavity field and the frequency of the excitons $\Delta_c^{x,y}$, $x-$polarized photon pair and $y-$polarized photon pair can be degenerate in energy. The spectrums of the emitted $x$, $y$ polarized photons, in the strong coupling regime, show
peaks at the frequencies $\omega_k\approx\omega_u-\omega^{\pm}_{x,y}$ and $\omega_{l/m}\approx\omega^{\pm}_{x,y}$, where $\omega^{\pm}_{x,y}=\frac{1}{2}\left(\omega^{x,y}_c+\omega_{x,y}\pm\sqrt{(\Delta_c^{x,y})^2+4g^2}\right)$ are the frequencies of the polariton states.
The polarization-entangled photon pairs can be generated by making the emitted
$x-$polarized and $y-$polarized  photon pairs degenerate. For the positive (negative) values of $\Delta_c^{x,y}$, the peaks in the spectrum corresponding to $\omega_k\approx\omega_u-\omega_{x,y}^+$ and $\omega_{l/m}\approx\omega_{x,y}^+$ ($\omega_k\approx\omega_u-\omega_{x,y}^-$ and $\omega_{l/m}\approx\omega_{x,y}^-$) are stronger and the probability of generating photons for these frequencies is increased. Therefore, a large probability of generating degenerate photon pairs can be achieved by overlapping these stronger peaks in the spectrum.
There are three possible coupling cases of interest that can do this. {\em Case 1}: by making both $x-$polariton states  and $y-$polariton states degenerate, $\omega_{x}^{\pm}=\omega_y^{\pm}$ which can be achieved with $\Delta_c^x=-\Delta_c^y=\delta_x$ (see Fig.2(a) and 2(c));
{\em Case 2:}  by making one of the $x-$polariton state degenerate to the other $y-$polariton state ($\omega_x^{-}=\omega_y^{+}$, see Figs.\,2(b) and 2(d); or $\omega_x^{+}=\omega_y^{-}$) when $\Delta_c^x$ and $\Delta_c^y$ are of opposite sign;
{\em Case 3:}  by making $\omega_x^{+}=\omega_y^{+}$ ($\omega_x^{-}=\omega_y^{-}$) when both $\Delta_c^x$ and $\Delta_c^y$ are positive (negative). Optimum entanglement is achieved from case-1 and case-2 above for $\Delta_c^x=-\Delta_c^y$, which
we example in Fig.\,2 and Fig.\,3.

We stress that our calculated spectra are drastically different
to those predicted previously using a dressed state picture, where
the latter uses simple Lorentzian line widths for each state \cite{johne}. Moreover, in the strong coupling regime,
the cavity-assisted generated photon pairs ($S_c$)  completely
dominates the spontaneously emitted photons ($S_r$), and by several orders of magnitude.
This effect is similar the the cavity-feeding process that occurs
for an off-resonant cavity mode \cite{usPRLSubmit}, where
the leaky cavity mode emission dominates the spectrum.
\begin{figure}[t!]
\centerline{\includegraphics[width=3.5in]{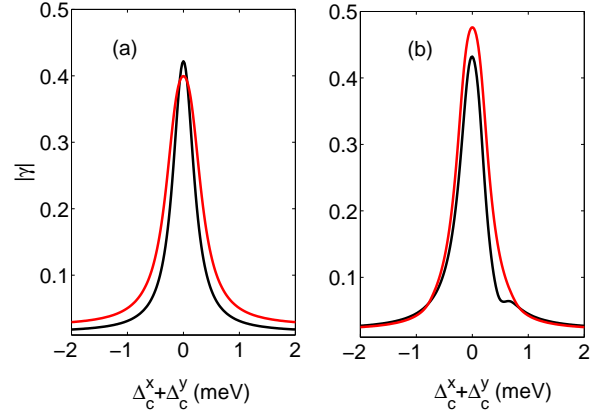}}
\vspace{-0.3cm}
\caption{
The amplitude of the off-diagonal element of the density matrix for filtered photon pairs, while keeping $\delta_x=0.1$\, meV fixed.
In (a) and (b) is shown the unfiltered and filter cases, respectively. The black curve represents $\Delta_c^x-\Delta_c^y=2\delta_{\rm x}$ (case 1), and  the red curve represents $\Delta_c^x-\Delta_c^y=-0.35\,$meV
(case 2); the other parameters are the same as in Fig.\,2.
The filter function corresponds to two spectral windows of width $w=0.2\,$meV centered at $\omega_x^{-}$ and $\omega_u-\omega_x^{-}$. Note that for $\Delta_c^x+\Delta_c^y=0$ corresponds to the optimal conditions for generating entangled photon pair as shown in Fig.\,(2).}
\vspace{-0.1cm}
\label{fig3}
\end{figure}

 The entanglement can be distilled by using frequency filters with a small spectral window $w$ centered at the frequencies of degenerate peaks in the spectrum of $x-$polarized and $y-$polarized photons. Subsequently, the response of spectral filter can be written as a projection operator of the following form
\begin{eqnarray}
W(\omega_k,\omega_l)=\left\{\begin{array}{cc}1,&{\rm for}
~|\omega_k-\omega_u+\omega^{\pm}_{x,y}|<w,\\1,&{\rm for}
~|\omega_l-\omega^{\pm}_{x,y}|< w,~~~~~~~~~\\
0,&{\rm otherwise.}~~~~~~~~~~~~~~~~~\end{array}\right.
\end{eqnarray}
After operating on the wave function of the emitted photons (\ref{state}), by spectral function $W(\omega_k,\omega_l)$ and tracing over the energy states \cite{akopian}, we get the reduced density matrix of the filtered photon pairs in the polarization basis. We consider the photon pairs in which one photon is emitted from the biexciton decay and the other is
emitted by the leaky cavity mode; in fact we can easily
neglect the spontaneous emission of both biexciton and exciton photons
as we have justified before.  The normalized off-diagonal element of the density matrix of photons is given by \cite{akopian}
\begin{equation}
\gamma=\frac{\int\int c^*_{5kl}(\infty)d_{5kl}(\infty)Wd\omega_kd\omega_l}{\int\int|c_{5kl}(\infty)|^2Wd\omega_kd\omega_l+\int\int|d_{5kl}(\infty)|^2Wd\omega_kd\omega_l}.
\end{equation}
The {\em concurrence}, which is a quantitative measure of entanglement, for the state of the filtered photon pair is given by $C=2|\gamma|$ \cite{dephasing}. The photons are thus maximally entangled when $|\gamma|=0.5$.
In Fig.\,(3), the value of $|\gamma|$ is plotted for two different cases of degenerate $x-$polarized and $y-$polarized photon pairs, corresponding to Figs.\,2(a) and 2(b); $\delta_x$ and $\Delta_c^x-\Delta_c^y$ are
fixed, while $\Delta_c^x+\Delta_c^y$ is changed, e.g., by
temperature of gas tuning; both unfiltered (a) and filtered (b) values are shown.
The spectral filter has negligible effect on case 1, but it improves the
concurrence of case 2 significantly. After filtering, the generated entangled photons, when both polariton states of $x-$polarized and $y-$ polarized photons are degenerate (in Fig.\,2(a)), have a smaller entanglement than the degenerate photons pairs
with an $\omega_x^-$ $x-$polarized polariton state and  an $\omega_y^+$ polariton$y-$ polarized state  (see Fig.\,2(b)). However, the photon source operating under the conditions of Fig.\,2(a) is a {\em deterministic entangled photon source}, while
the photon source operating under the conditions of Fig.\,2(b)--and using a spectral filter--is a
probabilistic photon sources,
as there is some probability of generating non-degenerate photon pairs. In both cases,  we get
significant concurrence values greater than 0.9.

Finally, we discuss the general criteria for achieving efficient entanglement. In general,
one desires to be in the strong coupling regime to overcome the
exchange splitting, thus the required conditions are  $g>\kappa$ and $g>\delta_{\rm x}/2$. To gain
insight into a smaller $g$ situation, we show in Fig.\,(4) the spectra
and entanglement that occurs for
$g=\kappa$ and for smaller values of $\delta_{\rm x}$. For the spectra (a-b), it is clear
that the indistinguishability of the $x-$polarized and
$y-$polarized pairs is increased, yet in (c-d) we see that
impressive entanglement values can still be achieved, even without a filter. In addition, use of a spectral filter can not improve entanglement significantly in these conditions.
Thus we believe that the general cavity-improvement could be significant
in the context of generated entangled photon pairs, and that thse
values are achievable using realistic experimentally parameters.
\begin{figure}[t!]
\centerline{\includegraphics[width=3.5in, height=3in]{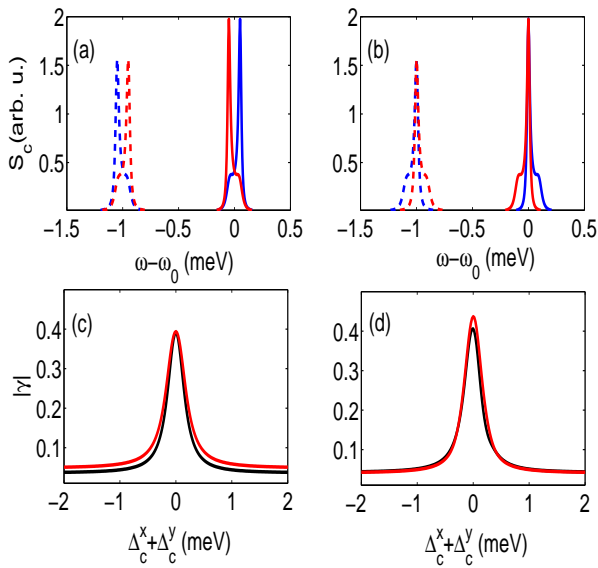}}
\vspace{-0.3cm}
\caption{
(a-b) As in in Fig.\,2(a-b), but with $g=\kappa=\delta_{\rm x}=0.05$ meV.
(c-d) As in Fig.\,3(a-b), but with $g=\kappa=\delta_{\rm x}=0.05$. The black curve represents $\Delta_c^x-\Delta_c^y=2\delta_{\rm x}$ (case 1), and  the red curve represents $\Delta_c^x-\Delta_c^y=-0.1\,$meV
(case 2). The filter function corresponds to two spectral windows of width $w=0.1\,$meV, centered at $\omega_x^{-}$ and $\omega_u-\omega_x^{-}$.}
\vspace{-0.1cm}
\label{fig4}
\end{figure}

\section{Conclusion}
In conclusion, we have derived
 and exploited analytical results for the wave functions of the emitted photon pairs from a QD embedded in a photonic crystal cavity that supports quasi-degenerate cavity modes.
In particular, we
have necessarily included finite exciton and biexciton level broadening, and damping of the leaky cavity mode,
and show that these relaxation mechanism must be included to connect to realistic
experiments. Finally, we have also discussed the method for optimizing and measuring the
entanglement between the emitted photons using a simple spectral filter.

\section{Acknowledgements}
We thank Robin Williams for useful discussions.
This work was supported by the National Sciences and Engineering Research Council of Canada,
and the Canadian Foundation for Innovation.

\end{document}